# Strain modulated band gap of edge passivated armchair graphene nanoribbons

## Xihong Peng, [1,]* Selina Velasquez[2]

[1] Department of Applied Sciences and Mathematics, Arizona State University, Mesa, AZ 85212

[2] College of Technology and Innovation, Arizona State University, Mesa, AZ 85212

## ABSTRACT

First principles calculations were performed to study strain effects on band gap of armchair graphene nanoribbons (AGNRs) with different edge passivation, including hydrogen, oxygen, and hydroxyl group. The band gap of the H-passivated AGNRs shows a nearly periodic zigzag variation under strain. For O and OH passivation, the zigzag patterns are significantly shifted by a modified quantum confinement due to the edges. In addition, the band gap of the O-passivated AGNRs experiences a direct-to-indirect transition with sufficient tensile strain (~ 5%). The indirect band gap reduces to zero with further increased strain, which may indicate a formation of metallic nanoribbons.

Keywords: armchair graphene nanoribbons, uniaxial strain, band structure, band gap, quantum confinement, edge passivation

Recently graphene, a two-dimensional (2D) sheet of $sp^2$-bonded carbon honeycomb lattice, has been considered as a promising material for many advanced applications in future electronics, such as ballistic single-electron transistors and interconnects.[1-3] The 2D graphene sheet demonstrates a zero band gap. For practical applications in semiconductor technology, the band gap of graphene has to be tuned to a finite value. A series of strategies were explored to engineer the band gap of graphene, for example, by applying an external electric field[4-7] or utilizing multilayer graphene structures.[7, 8] Tailoring the 2D graphene sheet into nanoribbons has been one of the promising approaches to create a finite value band gap. Individual factors, such as size[9-14], edge effect,[9, 15-17] and external strain,[18-23] can be employed to effectively tune the band gap of the graphene nanoribbons. However, it is still not clear what the combined effects of these factors are, especially strain and edge passivation, on the band gap of AGNRs

In present work, a theoretical study was conducted to investigate strain modulation of the band gap of the AGNRs with various edge passivation, including hydrogen, bridged oxygen and hydroxyl group. It was found that the zigzag pattern of strain-dependence of the band gap is significantly shifted by different passivation. In addition, a transition from direct to indirect band gap in the O-passivated AGNRs is observed by applying tensile strain around 5%. The ribbons could become metallic with further increased tensile strain.

Density-functional theory (DFT)[24] calculations were performed using VASP code.[25, 26] Local density approximation (LDA) was applied. In detail, a pseudo-potential plane wave approach was employed with a kinetic energy cutoff of 400.0 eV. Core electrons were described using Vanderbilt ultra-soft pseudo-potentials (US-PP).[27] Projector augmented wave (PAW) potentials[28, 29] were also used to check the calculations and no significant difference in the results was found between US-PP and PAW. Reciprocal space was sampled at 4 × 1× 1 using Monkhorst Pack meshes centered at Γ point. 21 K-points were included in band structure calculations. Dangling bonds on the edge of AGNRs were saturated in three scenarios: (1) by hydrogen atoms; (2) by oxygen atoms; and (3) by hydroxyl group (see Fig. 1). The initial lattice constant in a ribbon was set to be 4.22 Å, taken from the 2D graphene sheet. The lateral size of the simulation cell in the ribbon plane was chosen so that the vacuum distance between the ribbon and its replica (due to periodic boundary conditions) is more than 12 Å, and an 8 Å of vacuum separation was used to eliminate the interaction between ribbon layers. The total energy was converged to within 0.01 meV. Atoms were fully relaxed until forces are less than 0.02 eV/Å. The lattice constant along the armchair direction (i.e. *x*-axis) of all AGNRs was optimized through the technique of energy minimization.

The width L and the lattice constant *a* of a ribbon are defined as in Fig. 1(a). Based on the relaxed structure of a ribbon with an optimized lattice constant, uniaxial strain within the range of ±16% was applied by scaling the lattice constant (see Fig. 1(b)). The positive values of strain refer to uniaxial expansion, while negative corresponds to compression (note that the *y* and *z* coordinates of the ribbon are further relaxed at a given strain). It is known that, due to quantum confinement effects, AGNRs can be classified into three families according to the width L falling in the categories of 3n, 3n+1, and 3n+2, where n is a positive integer.[21, 23] In present work, AGNRs with a width of 12, 13, and 14 were chosen to represent those three families. In Table I, the studied AGNRs are listed with the relaxed lattice constants. It was found that the relaxed lattice constant varies with the edge passivation for a given width. The OH-passivated AGNR has the longest lattice constant, while the O-passivated ribbon has the shortest.

Strain effect on the band gap of the AGNRs is presented in Fig. 2. The band gap reported in the figure is measured at the Γ point. The band gap of the H-passivated AGNRs with different widths is plotted as a function of strain in Fig. 2(a). The graph shows a zigzag behavior with the maximum value of the band gap for the AGNRs with the width of 12, 13, and 14 occurring at +5%, -2%, and -7%, respectively, with the minimum value of the gap appearing at -6%, -10%, and +1%, respectively. The results are in a good agreement with literature.[22, 23] The zigzag patterns of the band gap with strain have been related to the movement of the Fermi point across discrete K-lines allowed by quantum confinement effects.[21, 23] Fig. 2(b) presents the band gap of the AGNRs of L = 13 with different edge passivation. Interestingly, the zigzag patterns of the O- and OH- passivated AGNRs are the same under negative strain, but shifted away from that of the H-passivated ribbon. Comparing them to Fig. 2(a), it was found that the O- and OH- passivated AGNRs with a width of 13 follows the zigzag pattern of the H-passivated AGNR with a width of 14. Fig. 2(c) shows the band gap of the O-passivated AGNRs with the widths of 12, 13, and 14. It was found that the O-passivated AGNRs with the widths 12, 13, and 14 demonstrate a similar zigzag behavior as the H-passivated ribbons of L = 13, 14, and 15, respectively. To illustrate this effect, Fig. 3(a) - 3(c) present the charge distributions of the valence band maximum (denoted by $v_1$) of the AGNRs with a width of 13, as an example. The pictures show that extra electron clouds contributed by oxygen atoms in the O- and OH-passivated AGNRs effectively extend the confined width of the nanoribbon, which may result in the observed shift.

In addition, with a detailed analysis of the band structures, it was found that the O-passivated AGNRs experience a direct-to-indirect gap transition with a sufficient tensile strain (~ 5%). This transition was not found in the H- and OH-passivated AGNRs within the range of strain considered in present work. As an example, the band structures of the H-, OH-, and O-passivated AGNRs with a width of 13 are presented in Fig. 4. Fig. 4(a) - 4(c) are the band structures for the H-passivated AGNR without and with strain ($\pm 16\%$). They all demonstrate a direct band gap at $\Gamma$. Similar results were found for the OH-passivated AGNR, shown in Fig. 4(d) - 4(f). However, the O-passivated AGNR displays a different behavior. From Fig. 4(g) and 4(h), the ribbon shows a direct band gap at $\Gamma$ for strain less than +5%. Within the strain range +5% to +10%, the band gap becomes indirect with the conduction band minimum located at the X point (see Fig. 4(i) and 4(j)). With +10% strain, the indirect band gap shrinks to zero. And with strain larger than +10%, no gap is observed, which may indicate a formation of a metallic AGNR. This dramatic change is originated from the strain-dependence of the two lowest conduction bands. For reference, the electronic states of these conduction bands were labeled using $c_1$ and $c_2$ at $\Gamma$, and $c_{1X}$ and $c_{2X}$ at X, where $c_{1X}/c_{2X}$ are degenerate. From Fig. 4(g) - 4(l), the energies of the $c_{1X}/c_{2X}$ states decrease with tensile strain. The nature of the band gap (direct or indirect) is determined by the lower energy of the electronic states $c_1$ and $c_{1X}/c_{2X}$.

To understand this transition of the band structure of the O-passivated AGNR, charge distributions were plotted for the electronic states $v_1$, $c_1$, $c_2$ and $c_{1X}/c_{2X}$ in Fig. 3(c) - 3(f). The electron clouds of $v_1$, $c_1$ and $c_2$ spread out in the ribbon, while the charge is highly localized on the edge atoms in the states of $c_{1X}/c_{2X}$. To illuminate the mechanism of the significantly decreased energies of the $c_{1X}/c_{2X}$ states under tensile strain, the structures of the relaxed and extremely strained ribbons (+16%) are presented in two adjacent simulation cells in Fig. 3(g) and 3(h). It shows that the tensile strain tears a carbon hexagon at the edge (formed by the carbon atoms labeled as 2, 4, 6, 1', 3', and 5' in Fig. 3(h)). The bond lengths between the edge atoms in the relaxed and strained ribbons are reported in Table II. For example, the bond lengths of the oxygen and the adjacent carbons (i. e. C1-O7 and C2-O7) are 1.51 Å in the relaxed AGNR, while they are 1.38 Å in the +16% strained ribbon. The bond lengths of C1-C3 and C2-C4 are 1.40 Å in the relaxed AGNR, while they are 1.37 Å in the strained ribbon. From Fig. 3(f), the charge in the $c_{1X}/c_{2X}$ states are primarily contributed by these four bonds. The reduction of the four bond lengths in the tensile strained ribbon make the electron cloud more effectively shared by the nuclei in the pentagon at the edge, this results in an appreciable decrease of the energy of the $c_{1X}/c_{2X}$ states due to an increased electron-nucleus attraction. Here, the difference in the electron-electron repulsion energy between the relaxed and strained ribbons is anticipated to be relatively small and the nucleus-nucleus interaction is taken as a constant shift in the total energy which is not included in the calculation of the electronic energies of the states.

In summary, it was found that (1) strain and edge passivation are alternative methods for tuning the band gap in the AGNRs; (2) the families of 3n, 3n+1 and 3n+2 of the O- and OH-passivated AGNRs demonstrate a similar zigzag behavior as the families of 3n+1, 3n+2, and 3(n+1) of the H-passivated AGNRs, respectively; (3) the band gap of the O-passivated AGNRs experiences a direct-to-indirect transition with sufficient tensile strain (~ 5%) and may display a metallic property.

This work is supported by the Research Initiative Fund from Arizona State University (ASU) to Peng. The authors thank the following for providing computational resources: ASU Fulton High Performance Computing Initiative (Saguaro) and National Center for Supercomputing Applications. Fu Tang and Paul Logan are acknowledged for the helpful discussions. Fu Tang is also greatly acknowledged and appreciated for the critical review of the manuscript.

* To whom correspondence should be addressed.  E-mail: xihong.peng@asu.edu.

**Table caption**

**Table I** **The studied AGNRs with the relaxed lattice constants. $N_C$, $N_H$, and $N_O$ represent the number of carbon, hydrogen and oxygen atoms in the unit cell, respectively.**

**Table II** **The bond lengths (in unit of Å) of the relaxed and +16% strained AGNR of L = 13 with O passivation. The number notation of atoms is indicated in Fig. 3(h).**

**Figure captions**

**Fig. 1 (Color online) The snapshots of AGNRs with a width of 13, passivated by hydrogen in (a); bridged oxygen in (c); hydroxyl group in (d); uniaxial strained in (b). Yellow, white, and red dots are C, H, and O atoms, respectively.**

**Fig. 2 (Color online) The DFT predicted band gap in AGNRs with different width and edge passivation as a function of uniaxial strain. The band gap is measured at the Γ point. Positive strain refers to uniaxial expansion while negative strain corresponds to its compression.**

**Fig. 3 (Color online) The charge density contour plots at iso-value 0.0004 for different states in the AGNR of L = 13 with (a) H, (b) OH, (c) to (f) O passivation. (g) and (f) The structures of the relaxed and +16% strained AGNR in two adjacent simulation cells.**

**Fig. 4 The band structures of the AGNR of L = 13 with different strain and edge passivation. The Fermi level is referenced at zero. The H- and OH- passivated AGNRs display a direct band gap at Γ. The O-passivated AGNR shows a direct gap at Γ with strain less than +5%. With strain in the range of +5% to +10%, the AGNR demonstrates an indirect band gap. With +10% strain, the indirect band gap shrinks to zero. Further increased tensile strain indicates a formation of metallic AGNR.**

| AGNRs | $N_C$ | $N_H$ | $N_O$ | Lattice $a$ (Å) |
|---|---|---|---|---|
| L12-H | 24 | 4 | 0 | 4.2529 |
| L12-O | 24 | 0 | 2 | 4.0705 |
| L12-OH | 24 | 4 | 4 | 4.2925 |
| L13-H | 26 | 4 | 0 | 4.2465 |
| L3-O | 26 | 0 | 2 | 4.0803 |
| L13-OH | 26 | 4 | 4 | 4.2837 |
| L14-H | 28 | 4 | 0 | 4.2453 |
| L14-O | 28 | 0 | 2 | 4.0996 |
| L14-OH | 28 | 4 | 4 | 4.2879 |

**Table I**

| Bond | Length for relaxed AGNR | Length for +16% strained AGNR |
|---|---|---|
| C1-O7 | 1.51 | 1.38 |
| C2-O7 | 1.51 | 1.38 |
| C1-C3 | 1.40 | 1.37 |
| C2-C4 | 1.40 | 1.37 |
| C3-C4 | 1.41 | 1.51 |
| C6-C5' | 1.36 | 1.53 |
| C4-C3' | 2.67 | 3.23 |
| C2-C1' | 1.59 | 2.47 |

**Table II**

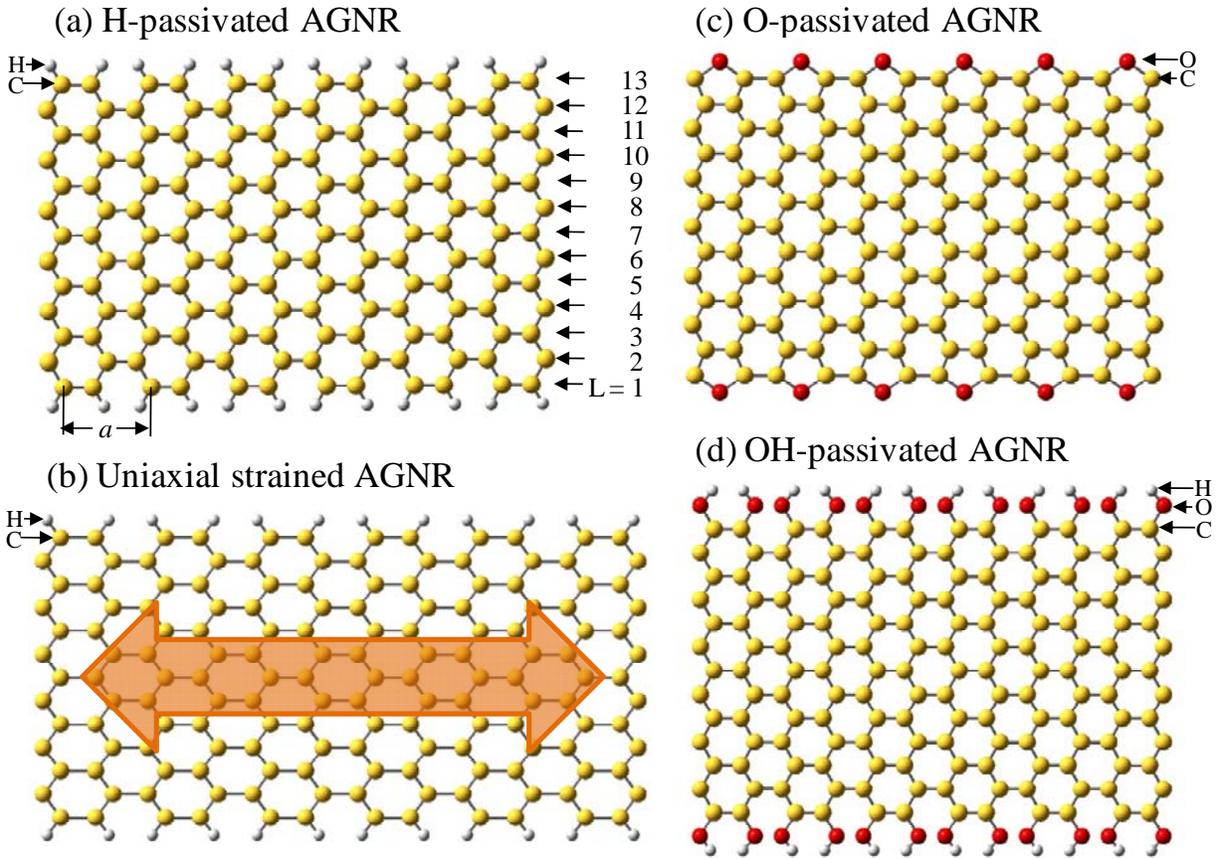

(a) H-passivated AGNR

(b) Uniaxial strained AGNR

(c) O-passivated AGNR

(d) OH-passivated AGNR

Figure 1

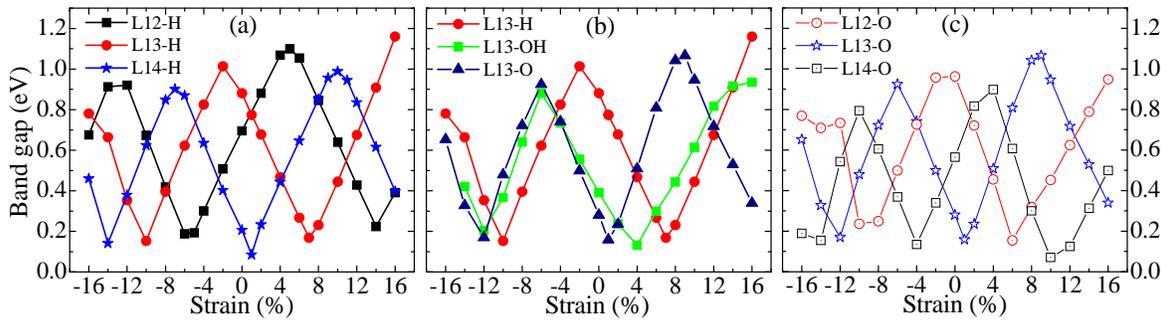

Figure 2

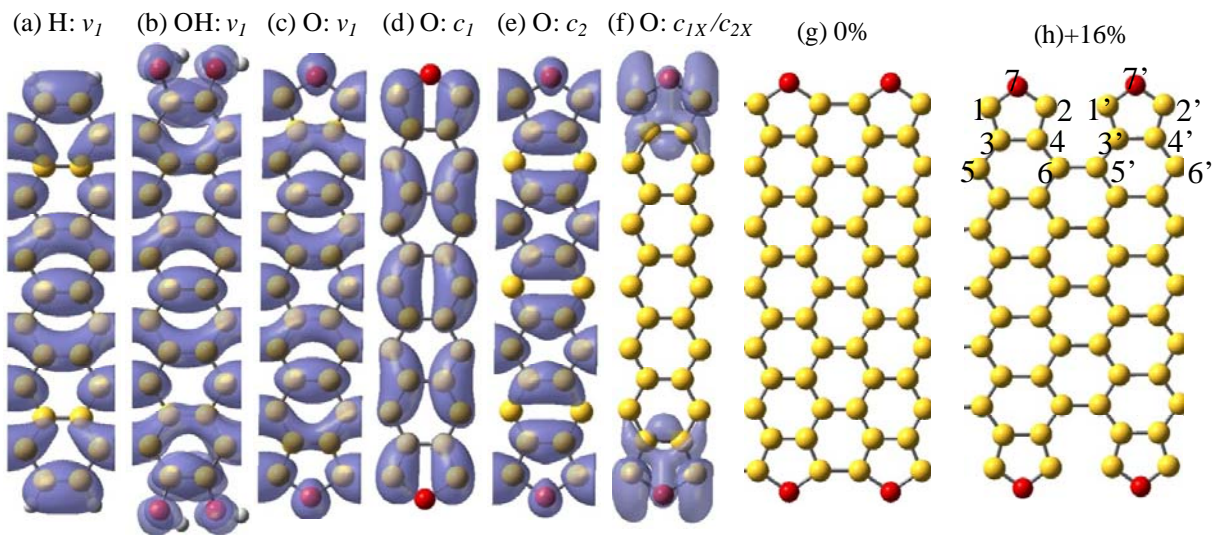

(a) H: $v_1$  (b) OH: $v_1$  (c) O: $v_1$  (d) O: $c_1$  (e) O: $c_2$  (f) O: $c_{1X}/c_{2X}$  (g) 0%  (h) +16%

Figure 3

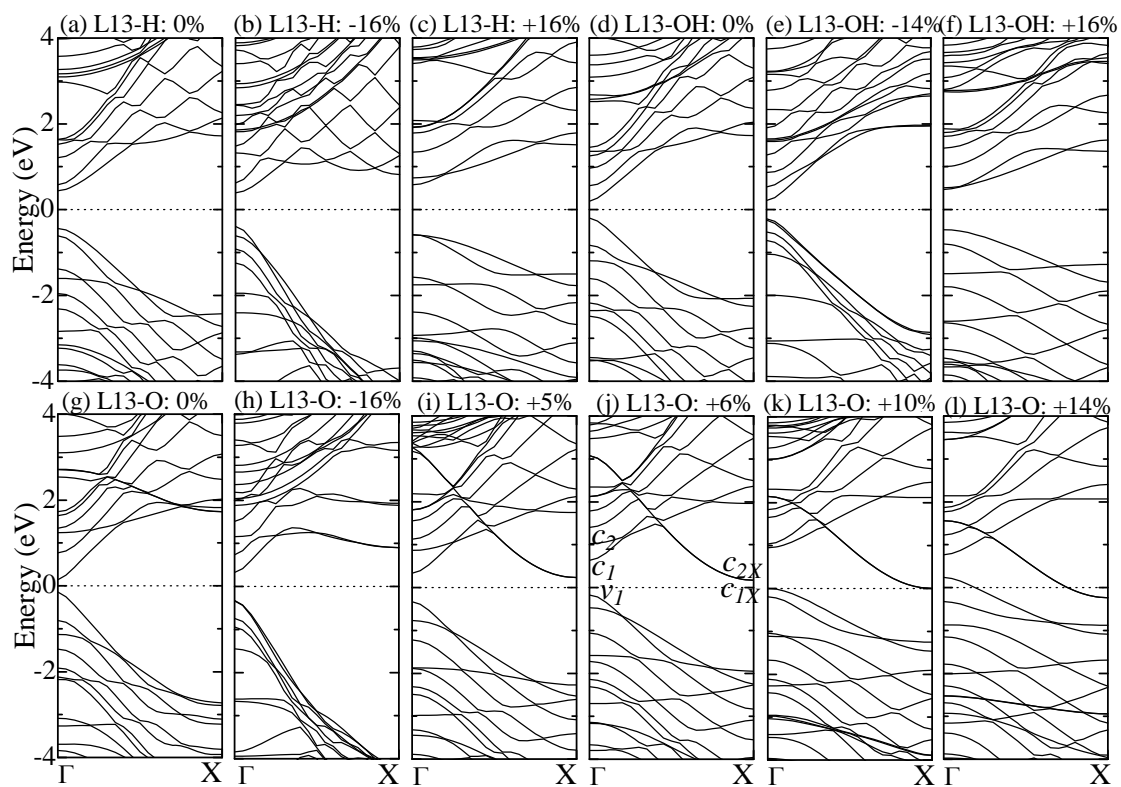

Figure 4